\documentclass[12pt]{article}
\usepackage{amsfonts}
\usepackage{epsfig,amsmath}
\usepackage{natbib}
\usepackage{amssymb}
\numberwithin{equation}{section}

\newcommand{\ex}[1]{\ensuremath{\mathbb{E}[#1]}}

\newcommand{\corr}[1]{\ensuremath{\mathrm{Corr}[#1]}}
\newcommand{\bd}[1]{\ensuremath{\mbox{\boldmath $#1$}}}

\begin{document}
\title{A Bayesian localised conditional auto-regressive model for estimating the health effects of air pollution}

\author{Duncan Lee$^{1,*}$,  Alastair Rushworth$^{1}$ and Sujit K. Sahu$^{2}$.\\
\vspace{0.5cm}\\
\small$^{1}$School of Mathematics and Statistics, University of Glasgow\normalsize\\
\small$^{2}$Southampton Statistical Sciences Research Institute, University of Southampton\normalsize} 

\maketitle

\begin{abstract}
Estimation of the long-term health effects of air pollution is a challenging task,  especially when modelling small-area disease incidence data in an ecological study design. The challenge comes from the unobserved underlying  spatial correlation structure in these data, which is accounted for  using random effects modelled by a globally smooth conditional autoregressive model. These smooth random effects confound the effects of air pollution, which are also globally smooth. To avoid this collinearity a Bayesian  localised conditional  autoregressive model is developed for the random effects. This localised model is flexible spatially, in the sense that it is not only able to model step changes in the random effects surface, but also is able to capture areas of spatial smoothness in the study region. This methodological development allows us to improve the estimation performance of the covariate effects, compared to using traditional conditional auto-regressive models. These results are established using a simulation study, and are then illustrated with our motivating study on air pollution and respiratory ill health in Greater Glasgow, Scotland in 2010. The model shows substantial health effects of particulate matter air pollution and income deprivation, whose effects have been consistently attenuated by the currently available globally smooth models. 
\\

\end{abstract}

\section{Introduction}

Quantification of the health effects of air pollution is an important statistical modelling problem that is of  considerable public interest, and  public health agencies and Government departments are required to estimate both its financial and health impact. For example,  in the UK the Department for the Environment, Food and Rural Affairs (DEFRA) estimate that \emph{``in 2008 air pollution in the form of anthropogenic particulate matter (PM) alone was estimated to reduce average life expectancy in the UK by six months. Thereby imposing an estimated equivalent health cost of $\pounds19$ billion''}, (\cite{defra2010}). These estimated effects are based on evidence from a large number of epidemiological studies, which have quantified the impact of both short-term and long-term exposure to air pollution. The effects of long-term exposure are  typically estimated from individual-level cohort studies such as \cite{hoek2002},  \cite{laden2006} and \cite{beverland2012}, but they are both expensive and time consuming to implement. Therefore, recent increases in the availability of small-area statistics has allowed these effects to be estimated using an ecological small-area design, with prominent examples being \cite{jerrett2005}, \cite{maheswaran2005}, \cite{elliot2007}, \cite{lee2009}, \cite{haining2010}, \cite{greven2011}  and \cite{lee2012b}. While these studies cannot assess the causal health effects of air pollution due to their ecological design, they are quick and cheap to implement, and they contribute to, and independently corroborate, the body of evidence about the long-term population level impact of air pollution.\\

The ecological design uses a geographical association study, where the study region of interest is partitioned into a set of non-overlapping areal units, such as counties, districts or postal codes. The number of disease cases arising from the population living in each areal unit is modelled, using Poisson regression, with a number of risk factors including average air pollution concentrations, measures of socio-economic deprivation and demography. Geographical (equivalently spatial) association is introduced into the model by means of a spatially correlated random effect for each areal unit.  These random effects model any residual spatial correlation that may be present in the disease data after the covariate effects have been removed, which may be caused by unmeasured confounding, neighbourhood effects (where individual areal unit's behaviour is influenced by that of neighbouring units) and grouping effects (where individual units seem to be close to similar units). These random effects, usually not of direct interest, are assigned a conditional autoregressive (CAR, \cite{besag1991}) prior as part of a hierarchical Bayesian model, for example see \cite{wakefield2007}. \\

The CAR model for spatial association tends to be globally smooth, and has only recently been shown (for example see \cite{reich2006}, \cite{hodges2010}, \cite{paciorek2010} and \cite{hughes2013}) to be potentially collinear with any covariate in the model which is also globally smooth, such as average air pollution concentrations.  Such collinearity leads to poor estimation performance for the fixed effects (\cite{clayton1993}), both in terms of the estimates and their associated levels of uncertainty. In addition, this collinearity suggests that the residual spatial correlation is unlikely to be globally spatially smooth, as that component of the spatial variation in the disease data will have been accounted for by the air pollution covariate. Instead, the residual spatial correlation can be strong in some areas showing smoothness, and can be weak in some other areas exhibiting  abrupt step changes. The existing CAR models force the random effects to exhibit a single global level of spatial smoothness determined by geographical adjacency, and are not flexible enough to capture the complex localised structure likely to be present in the residual spatial correlation.\\  

The lack of flexibility in existing CAR models and the collinearity problems highlighted by \cite{hodges2010} and others has motivated us to develop a new \emph{Localised Conditional AutoRegressive} (LCAR) prior for modelling residual spatial correlation, which is presented in Section~\ref{sn:method}. To contain the required flexibility the LCAR prior captures localised residual spatial correlation by allowing random effects in geographically adjacent areas to be correlated or conditionally independent, and we show that this prior distribution can have realisations at both spatial smoothing extremes, namely global smoothness and independence. 
This flexibility leads to a large increase in the computational burden and a lack of parsimony causing problems of parameter identifiability, and a critique of the limitations of the existing literature in this area is given in Section 2. \\

Here we solve these problems  with a novel prior elicitation method based on historical data, which is similar in spirit to power priors (see \cite{chen2006}). Our elicitation is based on an approximate Gaussian likelihood, and produces a set of candidate correlation structures for the residual spatial correlation. The LCAR prior thus combines a discrete uniform distribution on this set of candidate structures with a modified CAR prior for the random effects, which combined with the Poisson likelihood completes a full Bayesian hierarchical model. Inference is obtained using Markov chain Monte Carlo (MCMC) methods, and the model allows us to simultaneously estimate the random effects, their local spatial structure as well as the fixed effects. We conduct a large simulation study in Section~\ref{sn:simul} to show improved parameter estimation and model fitting when using the proposed LCAR prior distribution. The improvement, measured by the root mean square error (RMSE), is seen to be large for the fixed effects and somewhat substantial for model fitting. We follow up this investigation by analysing the motivating data set for the city of Glasgow in Section~\ref{sn:example}. But, first we present the motivating data set and discuss the background modelling and prior distributions in Section~\ref{sn:background}.



\section{Background}
\label{sn:background}

\subsection{Data}
\begin{figure}
\centering\caption{Maps displaying the spatial pattern in the standardised incidence ratio for respiratory disease (top panel) and the modelled yearly average concentration (in $\mu gm^{-3}$) of PM$_{10}$ (bottom panel).}\label{figure data1}
\begin{picture}(20,22)
\put(-2,0){\scalebox{1}{\includegraphics{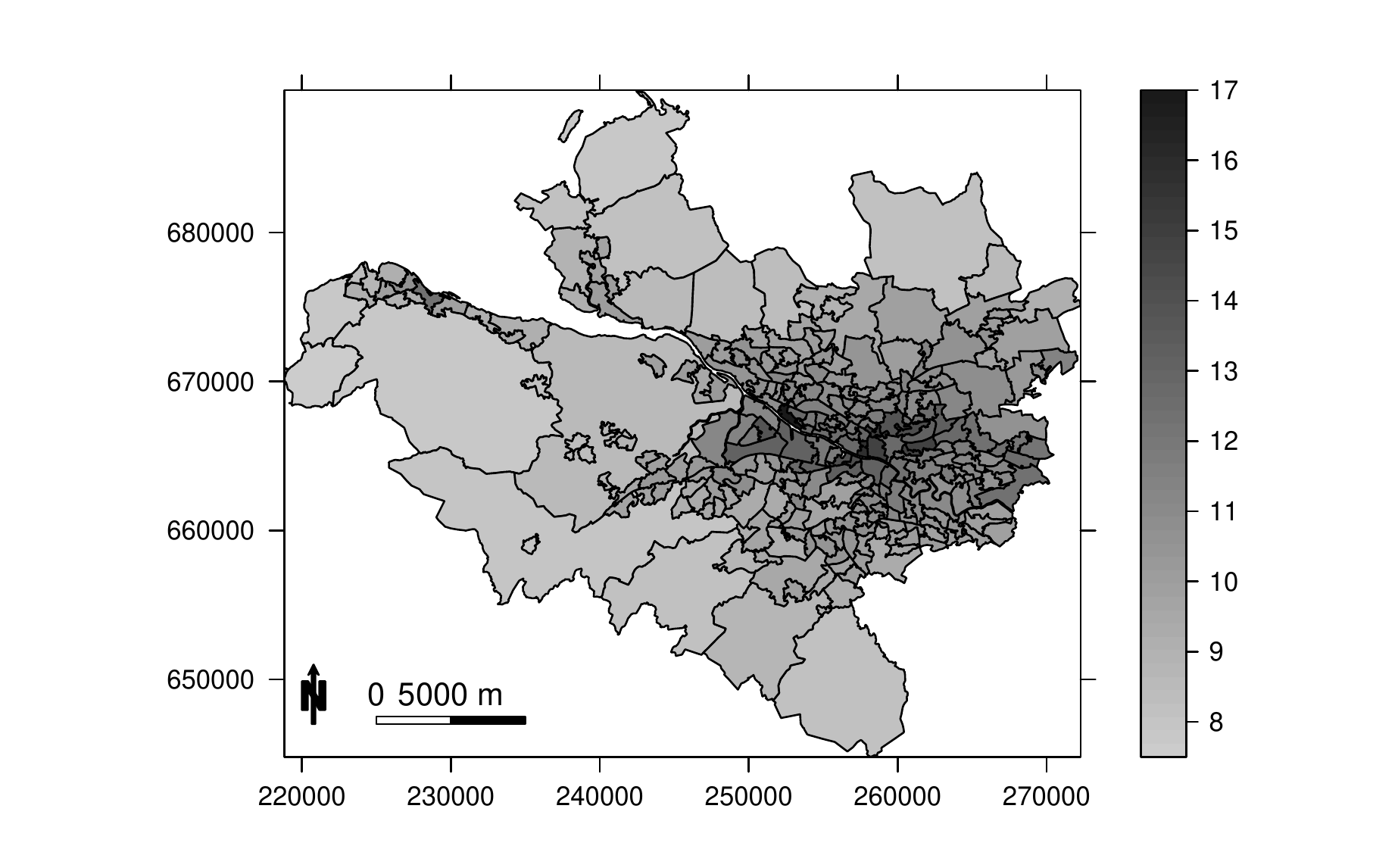}}}
\put(-2,11){\scalebox{1}{\includegraphics{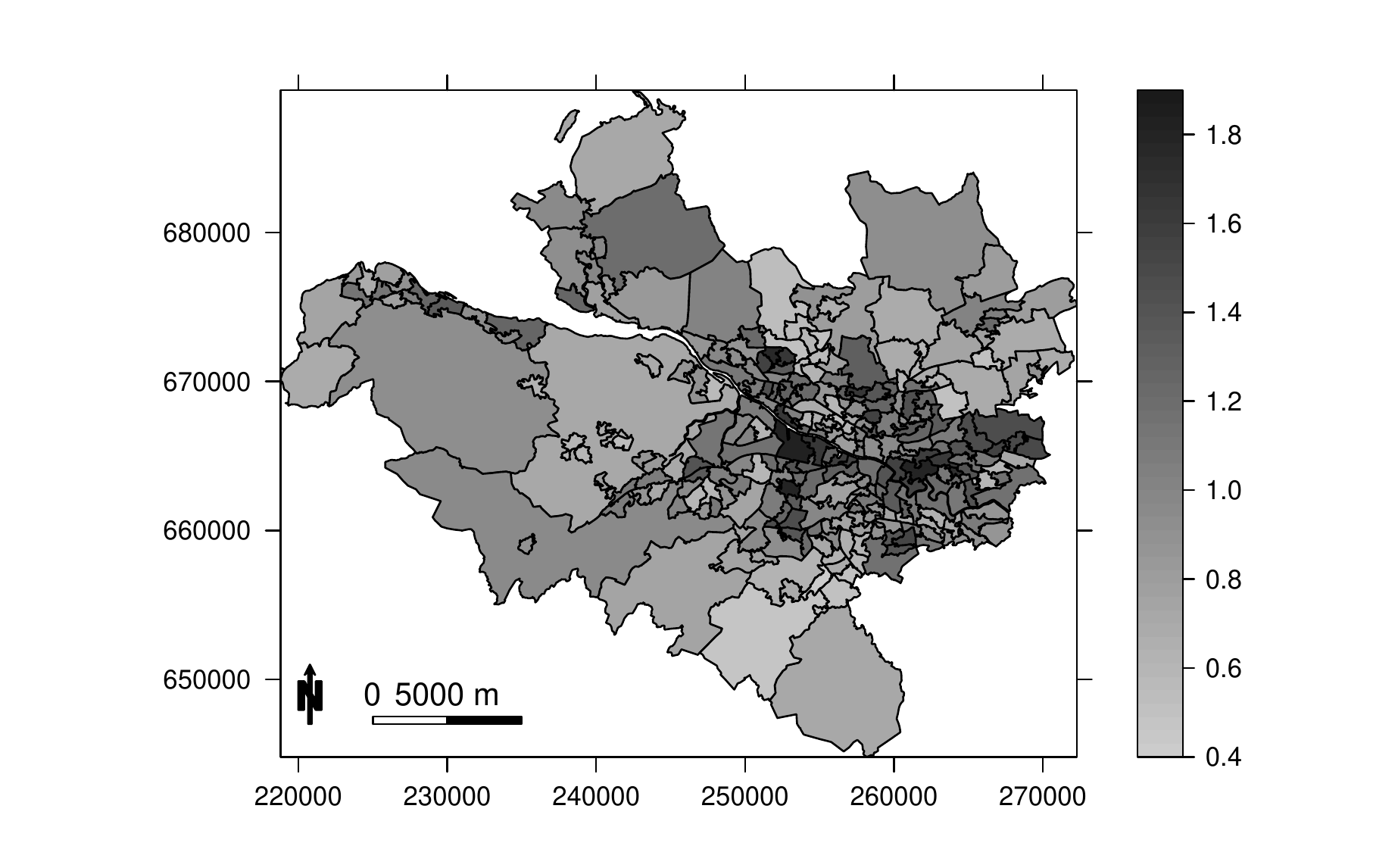}}}
\end{picture}
\end{figure}

The study region is the health board comprising the city of Glasgow and the river Clyde estuary,  which in 2010 contained just under 1.2 million people. The region is partitioned into $n=271$ administrative units called Intermediate Geographies (IG), which contain just over 4,000 people on average. The data used in this study are freely available, and can be downloaded from the Scottish Neighbourhood Statistics (SNS) database (\emph{http://www.sns.gov.uk}). The response variable is the numbers of admissions to non-psychiatric and non-obstetric hospitals in each IG in 2010 with a primary diagnosis of respiratory disease, which corresponds to codes J00-J99 and R09.1 of the International Classification of Disease tenth revision. Differences in the size and demographic structure of the populations living in each IG are accounted for by computing the expected numbers of hospital admissions using external standardisation, based on age and sex specific respiratory disease rates for the whole study region. An exploratory estimate of disease risk is given by the Standardised Incidence Ratio (SIR), which is the ratio of the observed to the expected numbers of admissions. It is displayed in the top panel of  Figure \ref{figure data1}, and shows that the risks are highest in the heavily deprived east end of Glasgow (east of the study region) as well as along the southern bank of the river Clyde, the latter of which flows into the sea in the west and runs south east through the study region.\\

Ambient air pollution concentrations are measured at a network of locations across Scotland, details of which are available at \emph{http://www.scottishairquality.co.uk/}. However, the network is not dense at the small-area scale required by this study, so instead we make use of modelled yearly average concentrations  at a resolution of 1 kilometre grid squares provided by the DEFRA (see \emph{http://laqm.defra.gov.uk/maps/}). We use concentrations for 2009 in this study rather than 2010, because it ensures that the air pollution exposure occurred before the hospital admissions due to respiratory illnesses. These modelled concentrations were computed using dispersion models and were then calibrated against the available monitoring data, and further details are available from \cite{grice2009}. They were subsequently converted to the intermediate geography scale by computing the median value within each IG. Concentrations (in $\mu gm^{-3}$) of nitrogen dioxide (NO$_{2}$) and particulate matter are available for this study, the latter being measured as both PM$_{10}$ (particles less than 10$\mu m$ in diameter) and PM$_{2.5}$ (particles less than 2.5$\mu m$ in diameter). The PM$_{10}$ data are displayed in the bottom panel of Figure \ref{figure data1}, which shows the highest concentrations are in the centre of the city of Glasgow as expected. \\

A number of other covariates were considered in this study, the most important of which is a measure of socio-economic deprivation. The relationship between deprivation and ill health is well known (for example see \cite{mackenbach1997}), and in this study we use the percentage of people living in each IG in 2009 who are defined to be income deprived, which means they are in receipt of a combination of means tested benefits. Other variables we also consider are measures of ethnicity (the percentage of school children in each IG who are non-white), access to alternative forms of health care (the average time taken to drive to a doctor's surgery) and a measure of urbanicity (a factor variable with 6 levels, with level one defined as urban and level six as rural).

\subsection{Modelling}
The study region is partitioned into $n$ areal units $\mathcal{A}=\{\mathcal{A}_{1},\ldots,\mathcal{A}_{n}\}$, and the vectors of observed and expected numbers of disease cases are denoted by  $\mathbf{Y}=(Y_{1},\ldots,Y_{n})$ and $\mathbf{E}=(E_{1},\ldots,E_{n})$ respectively. In addition, let $X=(\mathbf{x}_{1}^{\tiny\mbox{T}\normalsize},\ldots,\mathbf{x}_{n}^{\tiny\mbox{T}\normalsize})^{\tiny\mbox{T}\normalsize}$ denote the matrix of $p$ covariates, where the values relating to areal unit $\mathcal{A}_{k}$ are denoted by $\mathbf{x}_{k}^{\tiny\mbox{T}\normalsize}=(1, x_{k1}, \ldots, x_{kp})$. A Bayesian hierarchical model is typically used to model these data, and a general specification is given by

\begin{eqnarray}
Y_{k}|E_k,R_k &\sim&\mbox{Poisson}(E_kR_k)~~~~\mbox{for }k=1,\ldots,n,\nonumber\\
\ln(R_{k})&=&\mathbf{x}_{k}^{\tiny\mbox{T}\normalsize}\bd{\beta} + \phi_{k},\label{equation likelihood}
\end{eqnarray}
where the disease counts are assumed to be conditionally independent given the  covariates and the random effects. Here $\bd{\beta}=(\beta_0,\beta_1,\ldots,\beta_p)$ denotes the vector of covariate effects, while $R_k$ represents disease risk in areal unit $\mathcal{A}_k$. A value of $R_k$ greater (less) than one indicates that areal unit $\mathcal{A}_k$ has a higher (lower) than average disease risk, and in terms of interpretation,  $R_k=1.15$ corresponds to a 15$\%$ increased risk of disease. As previously discussed the random effects  $\bd{\phi}=(\phi_{1},\ldots,\phi_{n})$ capture  any residual spatial correlation present in the disease data, and are typically assigned a conditional autoregressive  prior, which is a special case of a Gaussian Markov Random Field (GMRF). Such models are typically specified as a set of $n$ univariate full conditional distributions, that is as $f(\phi_{k}|\bd{\phi}_{-k})$ for $k=1,\ldots,n$, where $\bd{\phi}_{-k}=(\phi_{1},\ldots,\phi_{k-1},\phi_{k+1},\ldots,\phi_{n})$. However, the Markov nature of these models means that the conditioning is only on the random effects in geographically adjacent areal units, which induces spatial correlation into $\bd{\phi}$. The adjacency information comes from a binary $n\times n$ neighbourhood matrix $W$, where $w_{ki}$ equals one if areal units $(\mathcal{A}_{k},\mathcal{A}_{i})$ share a common border (denoted $k\sim i$) and is zero otherwise (denoted $k\nsim i$). The intrinsic model (\cite{besag1991}, IAR) is the simplest prior in the CAR class, and its full conditional distributions are given by

\begin{equation}
\phi_{k}| \bd{\phi}_{-k}, \tau^{2}, W~\sim~\mbox{N}\left(\frac{\sum_{i=1}^{n}w_{ki}\phi_{i}}{\sum_{i=1}^{n}w_{ki}},~
\frac{\tau^{2}}{\sum_{i=1}^{n}w_{ki}}\right).\label{equation iar}
\end{equation}

The conditional expectation is the mean of the random effects in neighbouring areas, while the conditional variance is inversely proportional to the number of neighbours. The joint multivariate Gaussian distribution for $\bd{\phi}$ corresponding to (\ref{equation iar}) has a mean of zero but a singular precision matrix $Q(W)/\tau^{2}$, where $Q(W)=\mbox{diag}(W\mathbf{1}) - W$, and $\mathbf{1}$ is an $n$ dimensional vector of ones. This prior is appropriate if the residuals from the covariate component of the model, that is $\ln(\mathbf{Y}/\mathbf{E})-X\bd{\beta}$, are spatially smooth across the entire region, because the partial correlation between $(\phi_{k}, \phi_{j})$  conditional on the remaining random effects (denoted $\bd{\phi}_{-kj}$) is

\begin{equation}
\corr{\phi_{k},\phi_{j}|\bd{\phi}_{-kj},W}~=~\frac{w_{kj}}{\sqrt{(\sum_{i=1}^{n}w_{ki})(\sum_{i=1}^{n}w_{ji})}}\label{equation partialcorrelation}.
\end{equation}

Equation (\ref{equation partialcorrelation}) shows  that all pairs of  random effects relating to geographically adjacent areal units are partially correlated ($w_{kj}=1$), which smoothes the random effects across geographical borders. The most common extension to the IAR model to allow for varying levels of spatial smoothness is the BYM or convolution model (\cite{besag1991}), which augments the linear predictor in (\ref{equation likelihood}) with a second set of independent Gaussian random effects with a mean of zero and a constant variance. Further alternatives have been proposed by \cite{leroux1999} and \cite{stern1999}, but all of these extensions have a single spatial correlation parameter (for the BYM model it is the ratio of the two random effects variances) that controls the level of spatial smoothing globally across the entire region. Thus these models are inappropriate for capturing the likely localised nature of the residual spatial correlation, which may contain sub-region of spatial smoothness separated by step changes.\\ 

A small number of papers have extended the  class of CAR priors to account for localised spatial smoothing, the majority of which have treated $\mathcal{W}=\{w_{kj}|k\sim j, k>j\}$ as a set of binary random quantities, rather than forcing them to equal one. The neighbourhood matrix is always assumed to be symmetric so that changing $w_{kj}$ also changes $w_{jk}$, while the other elements in $W$ relating to non-neighbouring areal units remain fixed at zero. Equation (\ref{equation partialcorrelation}) shows that this allows $(\phi_{k},\phi_{j})$ corresponding to adjacent areal units to be conditionally independent or correlated, and if $w_{kj}$ (and hence $w_{jk}$) is estimated as zero  a boundary is said to exist between the two random effects. One of the first models in this vein was developed by \cite{lu2007}, who proposed a logistic regression model for the elements in $\mathcal{W}$, where the covariate was  a non-negative measure of the dissimilarity between areal units $(\mathcal{A}_{k},\mathcal{A}_{j})$. Similar approaches were proposed by \cite{ma2007} and \cite{ma2010}, who replace logistic regression with a second stage CAR prior and an Ising model respectively. However, these approaches introduce a large number of covariance parameters into the model, which for the Glasgow data considered here has $n = 271$ data points and $|\mathcal{W}|= 718$ spatial correlation parameters. Therefore, full estimation of $\mathcal{W}$ as a set of separate unknown parameters results in a highly overparameterised covariance model for $\bd{\phi}$, and \cite{li2011}  suggest that the individual elements are poorly identified from the data and are computationally expensive to update.\\

A related approach was proposed by \cite{lee2012}, who deterministically model the elements of $\mathcal{W}$ as a function of measures of dissimilarity  and a small number of parameters, rather than modelling each element as a separate random variable. However, their approach is designed for the related fields of disease mapping and Wombling (\cite{womble1951}), whose aims are not, as they are here, to estimate the effects of an exposure on a response.  An alternative approach was suggested by \cite{lee2013}, who propose an iterative algorithm in which $\mathcal{W}$ is updated deterministically based on the joint posterior distribution of the remaining model parameters. However, their algorithm has the drawback that only an estimate of each $w_{kj}$ is provided, rather than the posterior probability that $w_{kj}=1$.

\section{Methodology}
\label{sn:method}
Our methodological approach follows the literature critiqued above, and treats the elements in $\mathcal{W}$ relating to contiguous areal units as a set of binary random quantities. As conditional autoregressive priors are a special case of an undirected graphical model (\cite{lauritzen1996}), we follow the terminology in that literature and refer to $\mathcal{W}$ as the set of \emph{edges}, and further define any edge $w_{kj}\in\mathcal{W}$ that is estimated as zero as being removed. Our methodological innovation is a \emph{Localised Conditional AutoRegressive (LCAR)} prior, which comprises a joint distribution for an extended set of random effects $\tilde{\bd{\phi}}$ and the set of edges  $\mathcal{W}$, rather than the traditional approach of assuming the latter is fixed.  We decompose this joint prior distribution as $f(\tilde{\bd{\phi}}, \mathcal{W})=f(\tilde{\bd{\phi}}|\mathcal{W})f(\mathcal{W})$, and the next three sub-sections describe its two components as well as the overall hierarchical model.

\subsection{Prior distribution -  $f(\tilde{\bd{\phi}}|\mathcal{W})$}
The IAR prior given by (\ref{equation iar}) is an inappropriate model for $\bd{\phi}$ in the context of treating $\mathcal{W}$ as random, because of the possibility that all of the edges for a single areal unit could be removed. In this case $\sum_{i=1}^{n}w_{ki}=0$ for some $k$, resulting in (\ref{equation iar}) having an infinite mean and variance. Therefore we consider an extended vector of random effects $\tilde{\bd{\phi}}=(\bd{\phi},\phi_{*})$, where $\phi_{*}$ is a global random effect that is potentially common to all areal units and prevents any unit from having no edges. The extended $(n+1)\times (n+1)$ dimensional neighbourhood matrix corresponding to $\tilde{\bd{\phi}}$ is given by

\begin{equation}
\tilde{W} ~=~ \left[\begin{array}{ll}
W & \mathbf{w}_{*}\\ \mathbf{w}_{*}^{\tiny\mbox{T}\normalsize}& 0
\end{array}\right]\label{equation W matrix},
\end{equation}

where $\mathbf{w}_{*}=(w_{1*},\ldots,w_{n*})$ and $w_{k*}=\mbox{I}[\sum_{i\sim k}(1-w_{ki})>0]$. Here $\mbox{I}[.]$ denotes an indicator function, so that $w_{k*}=1$ if at least one edge relating to areal unit $\mathcal{A}_{k}$ has been removed, otherwise $w_{k*}$ equals zero. Based on this extended neighbourhood matrix we propose modelling $\tilde{\bd{\phi}}$ as $\tilde{\bd{\phi}}\sim\mbox{N}(\mathbf{0}, \tau^{2}Q(\tilde{W}, \epsilon)^{-1})$, where the precision matrix is  given by

\begin{equation}
Q(\tilde{W}, \epsilon) ~=~\mbox{diag}(\tilde{W}\mathbf{1}) - \tilde{W} + \epsilon I.\label{equation precision LCAR}
\end{equation}

The component $\mbox{diag}(\tilde{W}\mathbf{1}) - \tilde{W}$ corresponds to the IAR model applied to the extended random effects vector $\tilde{\bd{\phi}}$, while the addition of $\epsilon I$ ensures the precision matrix is diagonally dominant and hence invertible, with $\epsilon$ chosen to be a small positive constant. The requirement for $Q(\tilde{W},\epsilon)$ to be invertible comes from the need to calculate its determinant when updating $\mathcal{W}$, a difficulty not faced when implementing model (\ref{equation iar}) because $\mathcal{W}$ and $Q(W)$ are fixed. The addition of $\epsilon$ to the diagonal of the precision matrix  has been suggested in this context by \cite{lu2007}. The full conditional distributions corresponding to the LCAR model are given by:

\begin{eqnarray}
\phi_{k}|\tilde{\bd{\phi}}_{-k}&\sim&\mbox{N}\left(\frac{\sum_{i=1}^{n}w_{ki}\phi_{i} + w_{k*}\phi_{*}}{\sum_{i=1}^{n}w_{ki} + w_{k*} + \epsilon},~\frac{\tau^{2}}{\sum_{i=1}^{n}w_{ki} + w_{k*} + \epsilon}\right)\hspace{0.3cm} k=1,\ldots,n,\label{equation fc1}\\
\phi_{*}|\tilde{\bd{\phi}}_{-*}&\sim&\mbox{N}\left(\frac{\sum_{i=1}^{n}w_{i*}\phi_{i}}{\sum_{i=1}^{n}w_{i*} + \epsilon},~\frac{\tau^{2}}{\sum_{i=1}^{n}w_{i*} + \epsilon}\right).\nonumber
\end{eqnarray}

In (\ref{equation fc1}) the conditional expectation is a weighted average of the global random effect $\phi_{*}$ and the random effects in neighbouring areas, with the binary weights depending on the current value of $\mathcal{W}$. The conditional variance is approximately (due to $\epsilon$) inversely proportional to the number of edges remaining in the model, including the edge to the global random effect $\phi_{*}$. Removing the $kj$th edge from $\mathcal{W}$ sets $w_{kj}$ (and hence $w_{jk}$) equal to zero and makes $(\phi_{k}, \phi_{j})$ conditionally independent, and means that the global random effect $\phi_{*}$ is included in the conditional expectation to allow for non-spatial smoothing. In the extreme case of all edges being retained in the model (\ref{equation fc1}) simplifies to the IAR model for global spatial smoothing, while if all edges are removed the random effects are independent with mean and variance approximately (again due to $\epsilon$) equal to $\phi_{*}$ and $\tau^{2}$ respectively. 

\subsection{Prior distribution -  $f(\mathcal{W})$}
The dimensionality of $\mathcal{W}$ is  $N_{\mathcal{W}}=\mathbf{1}^TW\mathbf{1}/2$, and as each edge is binary the sample space  has size $2^{N_{\mathcal{W}}}$.  The simplest approach would be to assign each edge an independent Bernoulli prior, but as described in Section two this is likely to result in $\mathcal{W}$ being weakly identifiable. Therefore we treat $\mathcal{W}$ as a single random quantity, and propose the following discrete uniform prior for its neighbourhood matrix representation $\tilde{W}$;

\begin{equation}
\tilde{W}~\sim~\mbox{Discrete Uniform}(\tilde{W}^{(0)},\tilde{W}^{(1)},\dots,\tilde{W}^{(N_{\mathcal{W}})}).\label{equation DU prior}
\end{equation}

The last candidate value $\tilde{W}^{(N_{\mathcal{W}})}$  retains all $N_{\mathcal{W}}$ edges in the model, that is $w_{kj}=1~\forall~w_{kj}\in\mathcal{W}$, and corresponds to the IAR model for global spatial smoothing. Moving from $\tilde{W}^{(j)}$ to $\tilde{W}^{(j-1)}$ removes an edge from $\mathcal{W}$, which sets one additional $w_{kj}=w_{jk}=0$. This means that $\tilde{W}^{(0)}$ contains no edges and corresponds to independent  random effects. Thus the set $\{\tilde{W}^{(j)}|j=1,\ldots,N_{\mathcal{W}}-1\}$ corresponds to localised spatial smoothing, where some edges are present in the model and the corresponding random effects are smoothed, while other edges are absent and no such smoothing is enforced. This restriction reduces the sample space of $\mathcal{W}$ to being one-dimensional, because the possible values  $(\tilde{W}^{(0)},\tilde{W}^{(1)},\dots,\tilde{W}^{(N_{\mathcal{W}})})$ have a natural ordering in terms of the number of edges present in the model. \\

We propose eliciting  the set of candidate values $(\tilde{W}^{(0)},\tilde{W}^{(1)},\dots,\tilde{W}^{(N_{\mathcal{W}})})$ from disease data prior to the study period, because such data are typically available and should have a similar spatial structure to the response.  Let $((\mathbf{Y}_{1}^{p},\mathbf{E}_{1}^{p}),\ldots,(\mathbf{Y}_{r}^{p},\mathbf{E}_{r}^{p}))$ denote these vectors of observed and expected disease counts for the $r$ time periods prior to the study period. The general likelihood model (\ref{equation likelihood}) gives the vector of expectations for the study data as $\ex{\mathbf{Y}}=\mathbf{E}\exp(X\bd{\beta}+\bd{\phi})$, which is equivalent to $\ln\left(\ex{\mathbf{Y}}/\mathbf{E}\right)=X\bd{\beta}+\bd{\phi}$. Then as $\bd{\phi}\sim\mbox{N}(\mathbf{0}, \tau^{2}Q(\tilde{W}, \epsilon)_{1:n}^{-1})$, we make the approximation

\begin{equation}
\bd{\phi}_{j}^{p}~=~\ln\left[\frac{\mathbf{Y}_{j}^{p}}{\mathbf{E}_{j}^{p}}\right]~\approx~
\ln\left[\frac{\mathbf{Y}}{\mathbf{E}}\right]~\sim_{approx}~
\mbox{N}(X\bd{\beta}, \tau^{2}Q(\tilde{W}, \epsilon)_{1:n}^{-1})\hspace{1cm}\mbox{for }j=1,\ldots,r.\label{equation prior approximation}
\end{equation}

Based on this approximation the prior elicitation takes the form of an iterative algorithm, which begins at $\tilde{W}^{(N_{\mathcal{W}})}$ (which retains all edges in the model) and moves from $\tilde{W}^{(j)}$ to $\tilde{W}^{(j-1)}$ by removing a single edge from $\mathcal{W}$. The algorithm continues until it reaches $\tilde{W}^{(0)}$, where all edges have been removed. The algorithm moves from $\tilde{W}^{(j)}$ to $\tilde{W}^{(j-1)}$ by computing the joint approximate Gaussian log-likelihood for $(\bd{\phi}_{1}^{p},\ldots,\bd{\phi}_{r}^{p})$ based on  (\ref{equation prior approximation}). This is given by

\begin{eqnarray}
\ln[f(\bd{\phi}_{1}^{p},\ldots,\bd{\phi}_{r}^{p}|\tilde{W}^{(*)})]&=&\sum_{j=1}^{r}\ln[\mbox{N}(\bd{\phi}_{j}^{p}|X\hat{\bd{\beta}}, \hat{\tau}^{2}Q(\tilde{W}^{*}, \epsilon)_{1:n}^{-1})]\label{equation prior likelihood},\\
&\propto&\frac{r}{2}\ln(|Q(\tilde{W}^{*}, \epsilon)_{1:n}|) - \frac{nr}{2}\ln(\hat{\tau}^{2})\nonumber\\
&&-\frac{1}{2\hat{\tau}^{2}}\sum_{j=1}^{r}(\bd{\phi}_{j}^{p}-X\hat{\bd{\beta}})^{\tiny\mbox{T}\normalsize}Q(\tilde{W}^{*}, \epsilon)_{1:n}(\bd{\phi}_{j}^{p}-X\hat{\bd{\beta}}),\nonumber
\end{eqnarray}

and is calculated for all matrices $\tilde{W}^{(*)}$ that differ from $\tilde{W}^{(j)}$ by having one additional edge removed. From this set of candidates $\tilde{W}^{(j-1)}$ is equal to the value of $\tilde{W}^{(*)}$ that maximises the above log-likelihood. This prior elicitation approach removes edges from $\mathcal{W}$ in sequence conditional on the current value of $\mathcal{W}$, rather than naively treating each edge independently of the others. However, this approach requires (\ref{equation prior likelihood}) to be evaluated $N_{\mathcal{W}}(N_{\mathcal{W}}+1)/2$ times, which makes the approach computationally intensive. This computational burden is reduced by estimating $(\hat{\bd{\beta}}, \hat{\tau}^{2})$ by maximum likelihood, that is, based on $\tilde{W}^{(j)}$, $\hat{\bd{\beta}}=(X^{\tiny\mbox{T}\normalsize}Q(\tilde{W}^{(j)}, \epsilon)_{1:n}X)^{-1}X^{\tiny\mbox{T}\normalsize}Q(\tilde{W}^{(j)}, \epsilon)_{1:n}(\frac{1}{n}\sum_{j=1}^{r}\bd{\phi}_{j}^{p})$ and $\hat{\tau}^2=\frac{1}{nr}\sum_{j=1}^{r}(\bd{\phi}_{j}^{p}-X\hat{\bd{\beta}})^{\tiny\mbox{T}\normalsize}Q(\tilde{W}^{(j)}, \epsilon)_{1:n}(\bd{\phi}_{j}^{p}-X\hat{\bd{\beta}})$. In addition, to speed up the computation of the quadratic form in (\ref{equation prior likelihood}), the above estimators are based on $\tilde{W}^{(j)}$ rather than on each individual $\tilde{W}^{(*)}$.

\subsection{Overall model}
The Bayesian hierarchical model proposed here combines the likelihood (\ref{equation likelihood}) with the priors (\ref{equation fc1}) and (\ref{equation DU prior}) and is given by

\begin{eqnarray}
Y_{k}|E_k,R_k &\sim&\mbox{Poisson}(E_kR_k)~~~~\mbox{for }k=1,\ldots,n,\nonumber\\
\ln(R_{k})&=&\mathbf{x}_{k}^{\tiny\mbox{T}\normalsize}\bd{\beta} + \phi_{k},\label{equation full model}\\
\tilde{\bd{\phi}}&\sim&\mbox{N}(\mathbf{0}, \tau^{2}Q(\tilde{W}, \epsilon)^{-1}),\nonumber\\
\tilde{W}&\sim&\mbox{Discrete Uniform}(\tilde{W}^{(0)},\tilde{W}^{(1)},\dots,\tilde{W}^{(N_{\mathcal{W}})}),\nonumber\\
\beta_{j}&\sim&\mbox{N}(0, 1000)~~~~\mbox{for }j=1,\ldots,p,\nonumber\\
\tau^{2}&\sim&\mbox{Uniform}(0, 1000).\nonumber
\end{eqnarray}

Diffuse priors are specified for the regression parameters $\bd{\beta}$ and the variance parameter $\tau^{2}$, while $\epsilon$ is set equal to 0.001.  A sensitivity analysis to the latter is presented in Section four, which shows that model performance is not sensitive to this choice. Inference for this model is based on MCMC simulation, using a combination of Metropolis-Hastings and Gibbs sampling steps. The spatial structure matrix $\tilde{W}$ is updated using a  Metropolis-Hastings step, where if the current value in the Markov chain is $\tilde{W}^{(j)}$, then a new value is proposed uniformly from the set $(\tilde{W}^{(j-q)},\ldots,\tilde{W}^{(j-1)},\tilde{W}^{(j+1)},\dots,\tilde{W}^{(j+q)})$. Here $q$ is a tuning parameter, which controls the mixing and acceptance rates of the update. Functions to implement model (\ref{equation full model}) as well the prior elicitation are available in the statistical software \texttt{R} (\cite{R2009}), and are provided in the supplementary material accompanying this paper.

\section{Simulation study}
\label{sn:simul}

This section presents a simulation study comparing the relative estimation performances of the IAR, BYM and LCAR models, with regards to both fixed effects and fitted values.

\subsection{Data generation and study design}
\begin{figure}
\centering\caption{A map showing the piecewise constant mean function (with possible values $\{-1, 0, 1\}$) for the random effects that generate localised spatial correlation in the simulation study.}
\label{figure simstudy}\scalebox{0.5}{\includegraphics{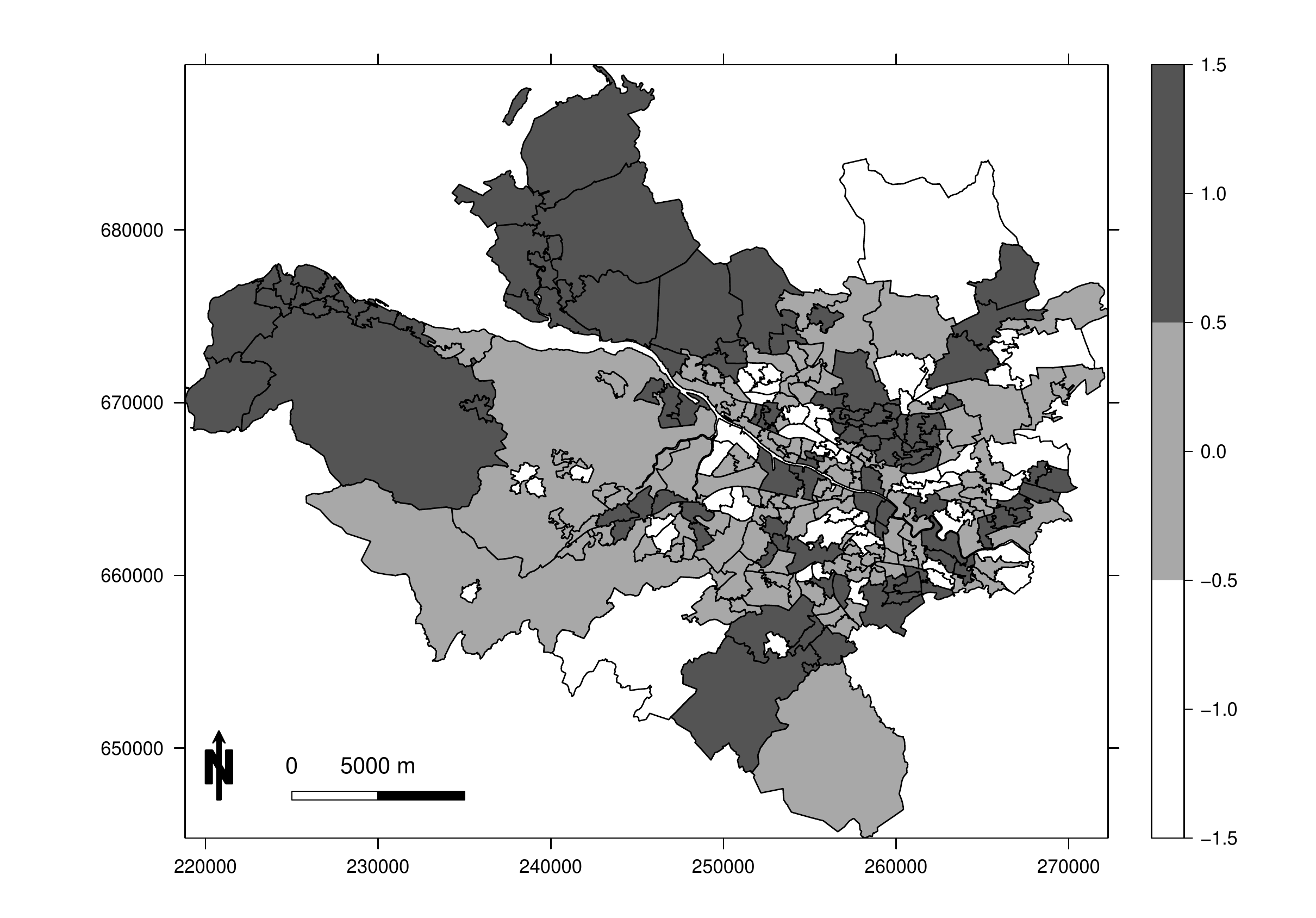}}
\end{figure}

Simulated data are generated for the $271$ Intermediate Geographies (IG) that comprise the Greater Glasgow and Clyde health board, which is the study region for the motivating study outlined in Section two. Disease counts are generated from model (\ref{equation likelihood}), where the size of the expected numbers $\mathbf{E}$ is varied to assess its impact on model performance. The log risk surface is generated from a linear combination of a single spatially smooth covariate and localised residual spatial correlation, with the former acting as the air pollution covariate while the latter will be modelled by the random effects. The pollution covariate is generated from a multivariate Gaussian distribution, where the mean is equal to zero and the  variance matrix is determined by a spatially smooth Mat\'{e}rn correlation function. The smoothness parameter equals 2.5 and the range parameter is chosen so that the median correlation between all pairs of areas is 0.5. The regression coefficient for this covariate is fixed at $\beta=0.1$, and is kept constant for each simulated data set. In contrast, new realisations of the covariate and the residual spatial correlation are generated for each simulated data set, to ensure the results are not affected by the particular realisations chosen. The residual correlation is also generated from a multivariate Gaussian distribution with a Mat\'{e}rn correlation function, where localised spatial structure is induced via a piecewise constant mean. The template for this is shown in Figure \ref{figure simstudy}, and only has three distinct values $\{-1, 0, 1\}$. These values are multiplied by a constant $M$ to obtain the expectation, where larger values of $M$ lead to bigger step changes in the spatial surface.\\

The study is split into nine different scenarios, which comprise all pairwise combinations of $M=0.5,1,1.5$ and each $E_{k}\in[10,25]$, $[50,100]$, $[150,250]$ for $k=1,\ldots,n$. The value of $\mathbf{E}$ specifies the underlying prevalence of the disease, while $M$ determines the extent to which the residual spatial correlation is locally rather than globally smooth, with larger values of $M$ corresponding to more prominent localised structure. Data generated under each scenario consists of the  study data and three years worth of prior data, which is the number of prior data sets used in the Glasgow motivating study. The residual spatial correlation for the latter is generated by adding uniform random noise in the range $[-0.1, 0.1]$ to  the realisation generated for the real data, which illustrates model performance in the realistic situation where the spatial patterns in the log risk surfaces for the prior and real data are similar but not identical.

\subsection{Results}
\begin{figure}
\centering\caption{Root mean square errors (RMSE) for the estimated regression parameter $\beta$ (top row) and the fitted values $E_kR_k$ (bottom row). In each case the dot represents the estimated RMSE while the black bars are bootstrapped 95$\%$ uncertainty intervals. The models are: (a) - IAR, (b) - BYM, (c) - LCAR.}\label{figure simulation1}
\begin{picture}(20,16.5)
\put(-2,0){\scalebox{0.85}{\includegraphics{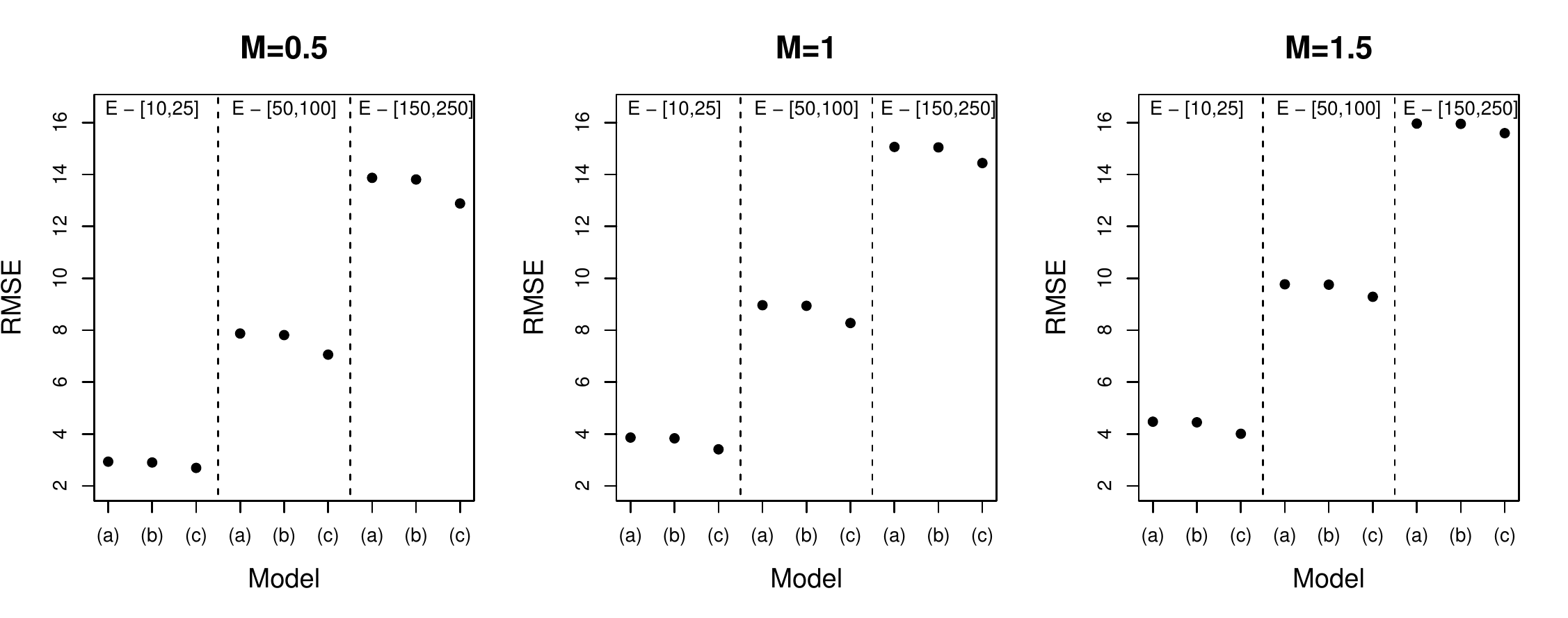}}}
\put(-2,8){\scalebox{0.85}{\includegraphics{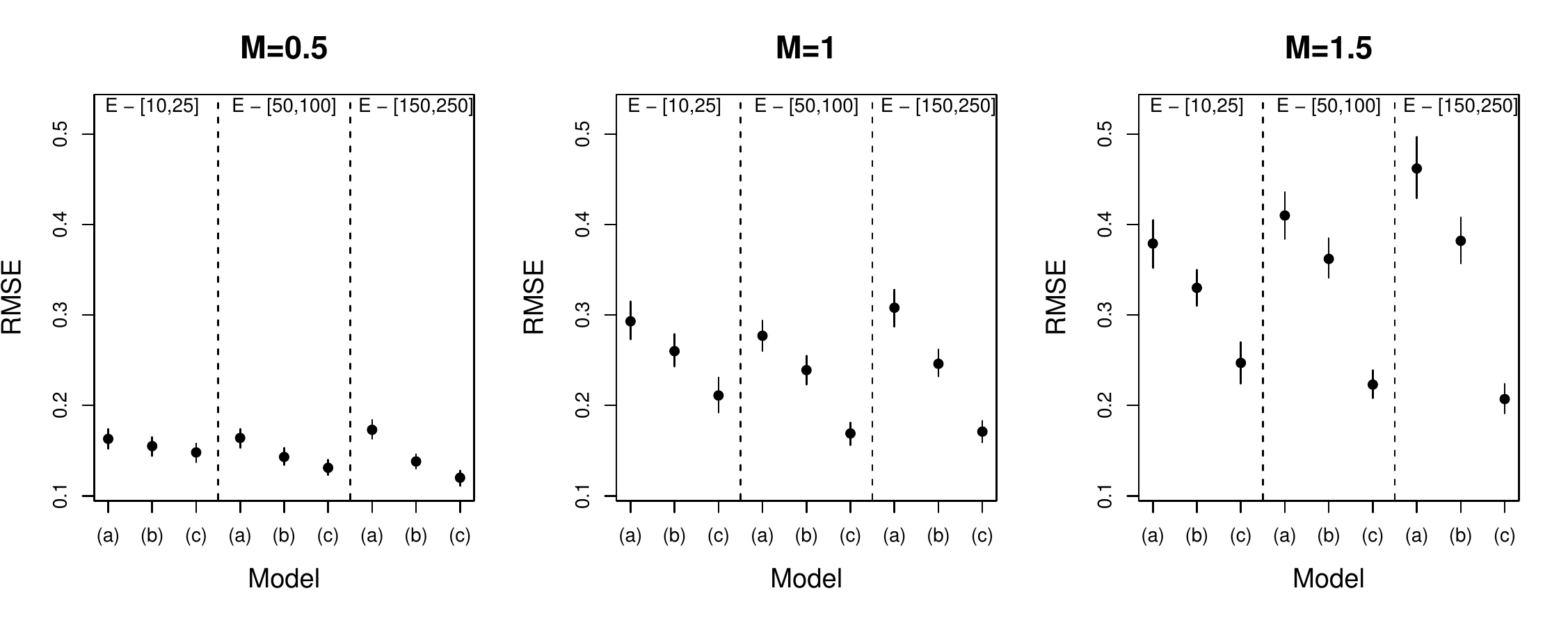}}}
\end{picture}
\end{figure}

Five hundred data sets are generated under each of the nine scenarios, and the results are displayed in Figure \ref{figure simulation1}. The back dots in the figure display the root mean square error (RMSE) values for all three models, for both the estimated regression parameter ($\beta$, top row) and the fitted values ($E_{k}R_{k}$, bottom row). The vertical lines represents bootstrapped 95$\%$ uncertainty intervals, which are based on 1000 bootstrapped samples. The main finding from the top row of the figure is that the RMSE for the regression parameter $\beta$ is always lowest for the LCAR model proposed here,  while the values for the IAR model are always highest. This is likely to be because the spatially smooth IAR model cannot represent the step changes present in the residual spatial correlation surface, where as the LCAR model is designed to do so. The results for the BYM model are always between these two extremes, which is unsurprising because while it can represent different levels of spatial smoothness, it effectively has a single parameter (the ratio of the two random effect variances) that controls the level of smoothing globally rather than locally as is required here. These results suggest that choosing a random effects model that can accurately capture the residual spatial structure is vital, as not doing so leads to vastly reduced estimation performance for the fixed effects.\\

In the present study the improved estimation performance for the LCAR model can be substantial, with percentage reductions in RMSE compared with the BYM model (the best competitor) ranging between 4.5$\%$ and 45.8$\%$. The figure also shows that the RMSE values from the LCAR model are mostly substantially better than those from the other two models, as the  uncertainty intervals do not overlap unless $M=0.5$, which is the case where the localised spatial structure is least prominent. In contrast, the greatest reductions in RMSE occur when $M=1.5$, which is the scenario in which the localised spatial structure in the residuals is most prominent. The estimation performance of all three models also reduces as $M$ increases, which is again likely to be because the localised nature of the residual spatial correlation becomes stronger. In contrast, changing the overall prevalence of the disease (i.e. changing $\mathbf{E}$) does not appear to have a large impact on the RMSE of $\beta$, and the direction of the small changes that are present are not consistent.\\

The bottom row of the figure shows that for the fitted values the LCAR model again always exhibits the best estimation performance, with lower RMSE values than the other three models in all cases. However, the differences are not as large as for $\beta$, with percentage reductions ranging between 2.2$\%$ and 11.1$\%$.  In addition, the bootstrapped uncertainty intervals in these estimates are very small, which is the reason they cannot be seen at the scale used in Figure \ref{figure simulation1}. Finally, a sensitivity analysis to the choice of the diagonally dominant constant $\epsilon$ was conducted, where the middle values of $M=1$ and $E_{k}\in[50,100])$ were used. Values of $\epsilon=0.0001, 0.001, 0.01$ were considered, and the results were robust to this choice.

\section{Results from the Glasgow study}
\label{sn:example}
 
\subsection{Modelling}
Initially, a  simple Poisson log-linear model including the four non-pollution covariates was fitted to the data, and only income deprivation exhibited a significant relationship with respiratory disease risk. The remaining three covariates were thus removed from the model, and each of the three pollution metrics (NO$_2$, PM$_{2.5}$ and PM$_{10}$) were added to income deprivation in separate models due to their collinearity. The residuals from these models exhibited substantial overdispersion, with estimates ranging between 3.05 and 3.10. The presence of residual spatial correlation was then assessed by performing a two-sided permutation test based on Moran's I statistic, which yielded highly significant p-values ranging between 0.0001 and 0.0002 depending on the pollutant included in the model. To alleviate these problems random effects were added to the model, and we compare the commonly used IAR and BYM specifications with the LCAR model proposed here. For the latter the prior elicitation was based on respiratory disease data from 2007 to 2009, the three years prior to the study period.

\subsection{Results - Model fit}
Posterior inference for all models was based on 3 parallel Markov chains, with a burn-in of 100,000 iterations, by which time convergence was assessed to have been reached, and then run for an additional 50,000 iterations, yielding 150,000 samples in total. The results are displayed in Table \ref{table application}, which quantifies the overall goodness of fit of the models and the estimated covariate effects. The results described in this section relate to models where PM$_{10}$ was the pollution metric, but the results for the other pollutants are similar. The goodness-of-fit of each model is summarised by its Deviance Information Criterion (DIC, \cite{spiegelhalter2002}), where a smaller value represents a better fitting model. The table shows that the LCAR model is the best fit to the data according to the DIC, with a value that is lower by 22.5 and 5.7 compared with the IAR and BYM models respectively. The presence of residual spatial correlation was then assessed using a two-sided Moran's I permutation test (based on 10,000 random permutations), and neither the LCAR nor the BYM models exhibited any remaining spatial structure. In contrast, the residuals from the IAR model exhibited substantial negative correlation, which is likely to be because the residual spatial structure from a covariate only model  is rougher than the spatially smooth random effects model. Therefore, as the residuals from the IAR model are essentially the difference between these two spatial surfaces, they are likely to have different signs in neighbouring areas, resulting in negative spatial correlation.\

\subsection{Results - covariate effects}
Table \ref{table application} also displays the estimated relationships between each covariate and the response, where all results are presented as relative risks for an increase of one standard deviation in each covariates value. The table shows that both particulate matter metrics exhibit substantial effects on respiratory disease risk, as apart from the IAR model, the worst performing model in terms of fixed effects estimation in the simulation study, the 95$\%$ credible intervals do not contain the null risk of one. The estimated relative risks for PM$_{10}$ across all three models range between 1.037 and 1.048 for a 1.5$\mu gm^{-3}$ increase in the yearly average concentration, while the corresponding risks for PM$_{2.5}$ range between 1.033 and 1.043 for a 1.1$\mu gm^{-3}$ increase. The lower ends of the credible intervals for NO$_{2}$ for all three models lie on the borderline of the null risk of one, with the IAR and BYM intervals including the null value while the interval for the LCAR model does not. The table also shows that income deprivation has a substantial effect on the response, with a 12.9$\%$ increase being associated with between a 32$\%$ and a 35$\%$ increased risk of respiratory disease. Consistent attenuation of the estimated covariate effects are observed for the IAR model compared with the other models, while the estimates from the BYM model are also consistently slightly lower than those from the LCAR model.

\begin{table}
\caption{A summary of the overall fit of each model (top panel) and the estimated covariate effects (bottom panel). The former includes the DIC and a permutation test for residual spatial correlation using Moran's I statistic. The latter are presented as relative risks for a one-standard deviation increase in each covariates value (in brackets).} \label{table application}
\centering\begin{tabular}{lrrr}
\hline
&\multicolumn{3}{c}{\textbf{Model}}\\
&\textbf{IAR}&\textbf{BYM} &\textbf{LCAR}\\\hline

DIC (p.d)&2113.3 (164.1)&2096.5 (164.6)&2090.8 (162.1)\\
Moran's I (p-value)&-0.0889 (0.013)& 0.0584 (0.111)& -0.0240 (0.519)\\\hline

Deprivation (12.9$\%$) &1.322 (1.286, 1.360)&1.343 (1.303, 1.381) &1.345 (1.311, 1.381)\\
NO$_{2}$ (5.0$\mu gm^{-3}$)&1.017 (0.975, 1.061)&1.032 (0.994, 1.067)&1.034 (1.002, 1.068)\\
PM$_{2.5}$ (1.1$\mu gm^{-3}$) &1.033 (0.990, 1.078)&1.042 (1.009, 1.078)&1.043 (1.010, 1.074)\\
PM$_{10}$ (1.5$\mu gm^{-3}$) &1.037 (0.997, 1.081)&1.043 (1.007, 1.079)&1.048 (1.017, 1.080)\\\hline
\end{tabular}
\end{table}

\subsection{Results - localised residual spatial correlation}
\begin{figure}
\centering\caption{Posterior density for the number of edges removed from the model. The three grey lines display the estimates from the individual Markov chains, while the bold black line displays the combined density from all three chains.}\label{figure W}
\scalebox{0.85}{\includegraphics{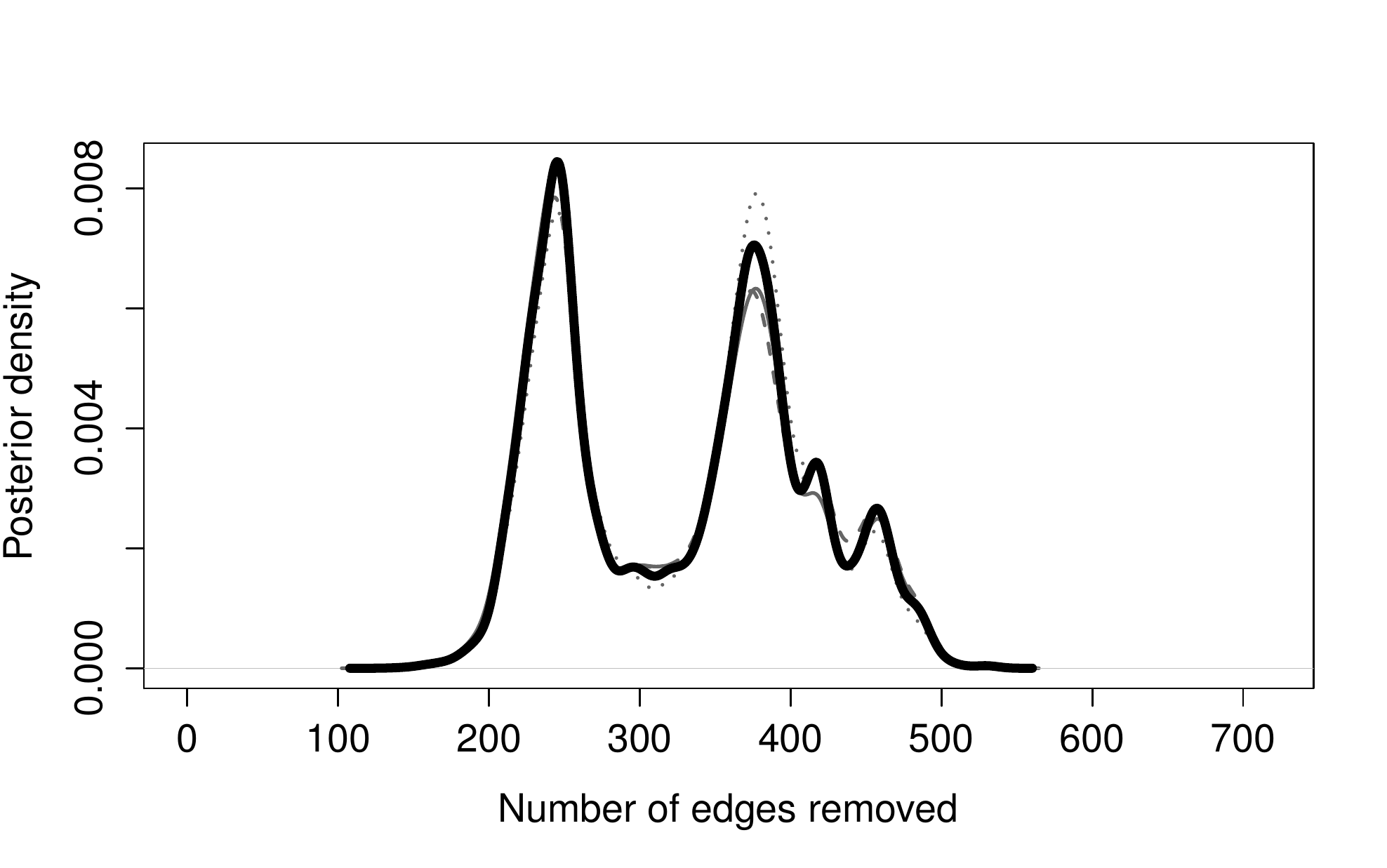}}
\end{figure}

Figure \ref{figure W} displays the posterior density for the number of edges removed from the model, where the three grey lines are chain specific estimates while the bold black line represents  the combined density from all three Markov chains. The figure shows close agreement between the chains, as all three give similar density estimates. There are 718 edges in total in the Greater Glasgow region, and the middle 95$\%$ of the posterior distribution lies between 210 and 473 of these having been removed. The figure suggests that while the posterior variability is relatively wide, there is information in the data to estimate the number of edges to remove. Specifically, the posterior distribution is multi-modal, with the two largest modes occurring when 245 and 375 edges are removed. The figure also provides strong evidence that the random effects are neither globally spatially smooth not independent, as there is no posterior mass at either end of the range of possible values (0 or 718 edges removed).

\section{Discussion}
This paper has proposed a new localised conditional autoregressive (LCAR) prior for modelling residual spatial correlation, which is flexible enough to capture either spatial smoothness or a distinct step change in the data between adjacent areal units. This flexibility is due to the treatment of the neighbourhood matrix $W$ as a random quantity to be estimated in the model, rather than assuming it is fixed based on geographical adjacency. However, this approach requires a large number of  covariance parameters to be estimated, and the resulting lack of parsimony is overcome by using prior information to greatly reduce the size and dimensionality of the sample space for $W$. The proposed model can estimate a range of localised spatial correlation structures, as well as patterns that are globally smooth or independent in space.\\

The simulation study has shown that the increased flexibility of the LCAR model results in superior estimation performance in a root measure error sense for both fixed effects and fitted values, when  compared with the commonly used global smoothing alternatives. This improvement in estimation is most prominent for covariate effects, where the percentage reductions in RMSE ranged between 4.5$\%$ and 45.8$\%$ in the study presented in Section four. The resulting conclusion to be drawn is that inappropriate control for residual spatial correlation can greatly retard fixed effects estimation, meaning that its careful modelling is vital even if it is not itself of direct interest.\\

The epidemiological study presented in this paper shows substantial evidence that particulate air pollution is harmful to respiratory health in Greater Glasgow, with an estimated increase in the population's disease burden of around 4$\%$  if yearly average concentrations increased by between 1.1 and 1.5 $\mu gm^{-3}$. However, one must remember that this is an observational ecological study design, and the results must not be interpreted in terms of individual level cause and effect (ecological bias). Even so, as small-area studies are cheaper and quicker to implement than individual level cohort studies, they form an important component of the evidence base quantifying the health effects of long-term exposure to air pollution.\\

There are many avenues for future work in this area, including the extension of the methodology to the spatio-temporal domain. In an epidemiological context the extension of the present study to the whole of the United Kingdom would be of interest to policymakers, as it would give the UK government a national rather than a regional picture of the extent of the air pollution problem. In addition, while the motivation for this paper was an ecological regression problem, the methodology developed will also be directly relevant to the fields of disease mapping (see for example \cite{lee2012}) and Wombling (\cite{womble1951}),  whose aims are to estimate the spatial pattern in disease risk and to identify any boundaries in the estimated risk surface.\\




\section*{Acknowledgements}
This work was funded by the Engineering and Physical Sciences Research Council (EPSRC) grant numbers EP/J017442/1 and EP/J017485/1, and the data and shapefiles were provided by DEFRA and the Scottish Government.\vspace*{-8pt}


\section*{Supplementary Materials}

The software (functions in R) to implement the LCAR model is available from the first author on request.\vspace*{-8pt}








\label{lastpage}

\end{document}